\documentclass[preprint,showpacs,preprintnumbers,amsmath,amssymb]{revtex4}

\usepackage{graphicx}
\usepackage{epstopdf}
\begin{document}

\title[The vanishing limit of the square-well fluid]{The vanishing limit of the square-well fluid: the adhesive hard sphere model as a reference system}

\author{  J.~Largo$^{1,2}$\footnote{Corresponding author. Present address: Dpt. de F\'{i}sica Aplicada, Universidad de Cantabria, Avd. Los Castros s/n Santander 39004, Spain.}, M.~A.~Miller$^3$, F.~Sciortino$^{1,2}$}
 \affiliation{ {$^1$Dipartimento di Fisica,
 $^2$INFM-CRS-SOFT, Universit\`a di Roma {\em La Sapienza}, Piazzale A. Moro  2, 00185 Roma, Italy} and
 $^3$ University Chemical Laboratory, Lensfield Road, Cambridge CB2 1EW, United Kingdom
   }
\affiliation{$^1$largojulio@gmail.com, $^3$mam1000@cam.ac.uk,$^2$francesco.sciortino@phys.uniroma1.it}

%
%\author{J.~Largo}
% \affiliation{ Dipartimento di Fisica,
% INFM-CRS-SOFT, Universit\`a di Roma {\em La Sapienza}, Piazzale A. Moro  2, 00185 Roma, Italy}
% \email{largojulio@gmail.com}
%\author{M.~A.~Miller}
% \affiliation{ University Chemical Laboratory, Lensfield Road, Cambridge CB2 1EW, United Kingdom}
% \email{mam1000@cam.ac.uk}
% \author{F.~Sciortino}
% \affiliation{ Dipartimento di Fisica,
% INFM-CRS-SOFT, Universit\`a di Roma {\em La Sapienza}, Piazzale A. Moro  2, 00185 Roma, Italy}
% \email{francesco.sciortino@phys.uniroma1.it}

\begin{abstract}
We  report a simulation study of the
gas-liquid critical point for the  square-well potential, for values of well width $\delta$ as small as 0.005 times the particle diameter $\sigma$.   For small $\delta$, the reduced second virial coefficient at the critical point  $B_2^{*c}$  is  found to depend linearly on $\delta$. The observed weak linear dependence is not sufficient to produce any significant observable effect if  the critical temperature $T_c$  is estimated via a constant  $B_2^{*c}$  assumption, 
due to the highly non linear transformation between $B_2^{*c}$ and $T_c$. This explains 
the previously observed validity of the law of corresponding states.  The critical density $\rho_c$  is also found to be constant  when measured in units of  the cubed average distance between two bonded particles $(1+0.5 \delta)\sigma$.   The possibility of describing the $\delta \rightarrow 0$ dependence with precise functional forms provides improved acccurate estimates of the critical parameters of the  adhesive hard-sphere AHS model.
\end{abstract}


\maketitle

\section{Introduction}
Investigation of protein and colloidal systems 
has focused  the attention of the scientific community on the phase-diagram behavior of short-range attractive potentials and of the role of the  range of interaction in controlling the thermodynamic and dynamic properties of the system\cite{Lekkerkerker,Pellicane2004,Rosenbaum1996,Lomakin1996, Poon1997}.  
Colloidal particles, due to their nano or microscopic size are often characterized by effective interactions\cite{likos-review} whose range is significantly smaller than the particle diameter. Under these conditions, it has been argued that the actual shape of the potential is irrelevant and that the thermodynamics\cite{Noro2000}, as well as the dynamics\cite{Foffi2005}, of  different systems approximately satisfy an extended law of corresponding states\cite{VliegenthartLekkerkerker2000}.  This law allows for a  comparison between different  systems, once the effective  diameter of the particle is known (i.e., when the repulsive part of the interaction can be  mapped into an equivalent hard-sphere diameter~\cite{Barker1967}). It has been proposed that  the  second virial coefficient, normalized by 
the corresponding hard-sphere second virial coefficient, $B_2^*$
may act as a proper scaling variable. Therefore systems with equal second virial coefficient and effective diameter should have similar thermodynamical properties.

The adhesive hard-sphere (AHS) potential\cite{Baxter1968}, the limiting behavior  of an infinitesimal interaction range coupled to infinite interaction strength such that $B_2$ is finite, has also received significant attention.
For this potential $B_2^*=1-1/4\tau$, where $\tau$, which acts as an effective scaled temperature, is the so-called stickiness parameter.  Despite the known thermodynamic anomalies\cite{Stell1991},  an analytic evaluation of the (metastable) critical point within the 
Percus-Yevick closure with both the energy and the compressibility routes for this potential is available\cite{Watts1971,Baxter1968}.   
In the energy route  the  critical
point is located at $B_2^{*c}=-1.1097$ ($\tau_c=0.1185$) and number density $\rho=0.609$.

The availability of analytic predictions for this model has favored its application in the interpretation of  experimental data for several disparate colloid (and protein) systems\cite{TartagliaJPCM,verduin,CaccamoPhysRep,Pellicane2004,Rosenbaum1996,Lomakin1996, Poon1997},
an application whose validity has been reinforced by the
extended law of corresponding states.    For this reason, it
is important to try to accurately estimate the properties of the AHS model as a reference, to support existing predictions or to suggest improvements to available theoretical approaches. Numerical simulations of the AHS model
have been attempted in the past\cite{seatonglandt1987, kranendonkfrenkel,Lee2001}.
 A recent effort in  the direction of evaluating the phase diagram of the model has been provided by Miller and Frenkel\cite{Miller2003}, based on an ingenious identification of the appropriate Monte Carlo (MC) moves for this
 potential\cite{seatonglandt1987, kranendonkfrenkel}.  Their study
provides an estimate of the location of the critical point 
at $B_2^{*c}=-1.21(1)$ ($\tau_c=0.1133(5)$) and  $\rho_c=0.508(10)$.

In this article we propose a different approach to the
numerical evaluation of the critical properties of the
AHS model, based on extrapolation of standard grand
canonical MC simulation results for a sequence of square well (SW)
potentials with progressively smaller attraction ranges $\delta$ 
(down to  $\delta=0.005$, in units of the  hard-sphere diameter $\sigma$).

The SW potential is defined as:
\begin{equation}
\label{eq:SWpotential}
U(r) = \left\{ 
\begin{array}{ll} 
\infty & \textrm{if $r\leq\sigma$}\\ 
 -\epsilon & \textrm{if $\sigma < r\leq\sigma+\delta \sigma$}\\ 
0 & \textrm{if $r>\sigma+\delta \sigma$} 
\end{array} \right.
\end{equation}
\noindent
The SW fluid has been profusely studied \cite{Vega1992,DelRio2002,LopezRendon2006,pagangunton,liukumar, Elliot1999,Chang1994} for $\delta>0.1$. It has been shown\cite{pagangunton,liukumar} that for $\delta\lesssim 0.25$ 
gas-liquid separation becomes metastable with respect to
the fluid-solid equilibrium.  Despite its metastable character,
investigation of smaller $\delta$ values retains its importance, since the crystallization time is often much longer than the
experimental one and
gas-liquid phase separation is readily accessed (an effect facilitated by the 
large difference between the fluid density and the crystal density and, in experiments, by the intrinsic sample polydispersity) .

%Recently some authors~\cite{LopezRendon2006} argue  the breakdown of  the law of corresponding states at criticality takes place for $\delta$ values greater than 0.125. Another interesting article~\cite{Malijevsky2006} has studied the equivalence between the AHS fluid and the  SW fluid as the $\delta$ values are reduced from a structural point of view CHIARIRE. It seems that the accurate determination of the correct limit of application of the extended corresponding states law is missing. 
 
Despite the importance of the SW model in relation to
the AHS potential, 
no studies of the  $\delta$-dependence of the critical point location  has been previously  reported for very small 
  $\delta$.    This is in large part due to the fact that for smaller  and smaller $\delta$, the critical temperature significantly decreases.  Indeed, according to the constant $B_2^{*c}$ prediction of Noro and Frenkel\cite{Noro2000}, it should vary as
\begin{equation}
\label{eq:Tc}
\frac{k_B T_c}{\epsilon}=\left[\ln \left (1+\frac{1-B_2^{*c}}{(1+\delta)^3-1}\right )\right]^{-1}
\end{equation}
\noindent
  making bond-breaking (changes of the particle energy of order $\epsilon$) events rarer and rarer in the simulation.
Moreover, the size of the translational step in the MC code  is of the order of  $\delta$.  On the other hand, 
     the location of the  critical point becomes  more and more metastable which, in principle, poses a limit to the smallest 
     $\delta$ which can be studied. 
 
Despite these numerical difficulties, we have been able
to estimate the location of the critical point down to 
$\delta=0.005 \sigma$. We present here the $\delta$ dependence of
the critical temperature and density and the values of the  second virial coefficient and energy at the critical point. 
In all cases, a short linear extrapolation to $\delta=0$  provides novel accurate estimates of the corresponding quantities for the AHS model.

\section{Monte Carlo simulations}

We have simulated the SW system in the grand canonical (GC) ensemble in order to locate the gas-liquid  critical point for different $\delta$ values. The critical point is identified by mapping the grand canonical density distribution onto the universal Ising model distribution, following the method described by Bruce and Wilding\cite{Bruce1992}.  Histogram rewighting \cite{Wilding1997} was used to achieve an accurate estimate of the critical point, and the field mixing parameter was always found to be negligible.
We define a Monte Carlo step as one hundred trial moves, with 
an average of 95\% translation and 5\% trial insertions or removals  of one particle in the system. A translational move is defined as
a displacement in a random direction by a random amount between $\pm \delta/2$. We simulate different realizations of the same system (at the same chemical potential
and temperature) to improve statistics. Simulations
lasted more than $10^7$  MC steps.  We  have studied 
cubic boxes of side 5$\sigma$ and/or 8$\sigma$, to estimate the importance of finite-size effects. 
Additionally, for $\delta=0.05$ we have studied several other box sizes
to  estimate the deviation of the value of the critical parameter 
 for the bulk limit case.  Proper finite size studies for  $\delta<0.05$ are at this moment numerically prohibitive.  Histogram reweighting was used to map the density distribution of the fluid onto the universal critical order parameter distribution of the Ising model\cite{Wilding1997}, thereby reaching an accurate estimate of the critical point.   The field mixing parameter was always been found to be negligible. 
  
\begin{figure}[htbp]
\begin{center}
\includegraphics[width=15 cm, clip=true]{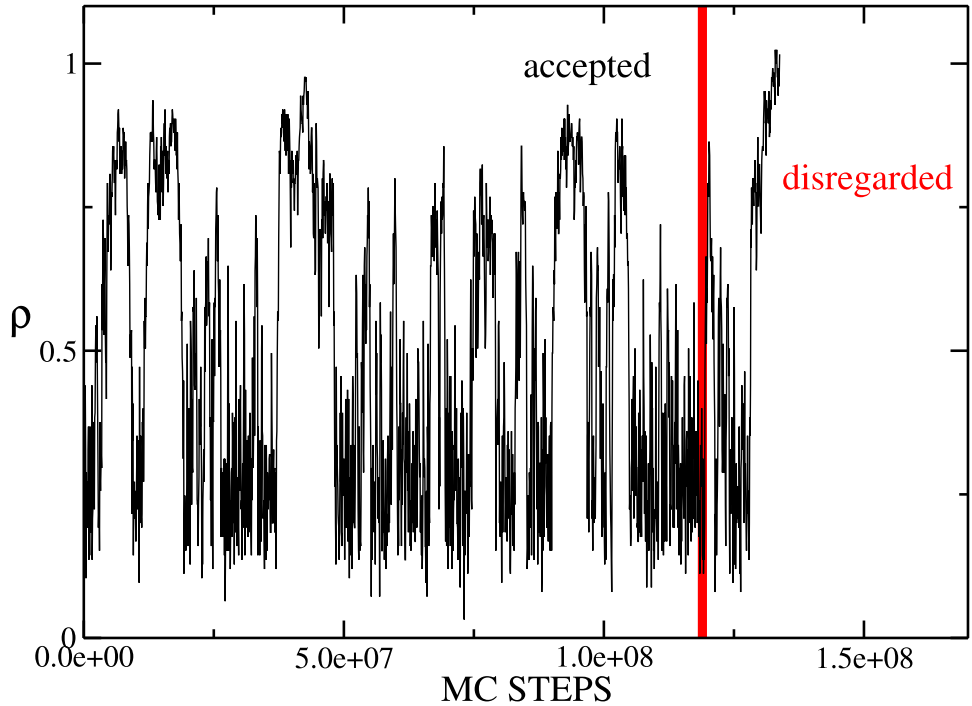}
\caption{Density evolution for a single run of a SW fluid with $\delta$=0.01, $T^{*}=0.233$, and $\mu/kT=-2.41$ (close to the critical point).  At long time,  a transition to a
dense phase is observed and, consequently the simulation on the right side of the vertical line
is disregarded.}
\label{fig:densevol}
\end{center}
\end{figure}

We have occasionally observed a transition to a more dense stable phase, signaling that indeed the
values of the chemical potential studied admit metastable fluid solutions.  In all cases where a transition to a
more dense phase was observed, the simulation was interrupted and the 10\% of configurations saved just before the transition were disregarded.
This procedure is shown graphically in Fig.-\ref{fig:densevol}.

\section{Results}

\begin{figure}[htbp]
    \centering
    \includegraphics[width=14 cm, clip=true]{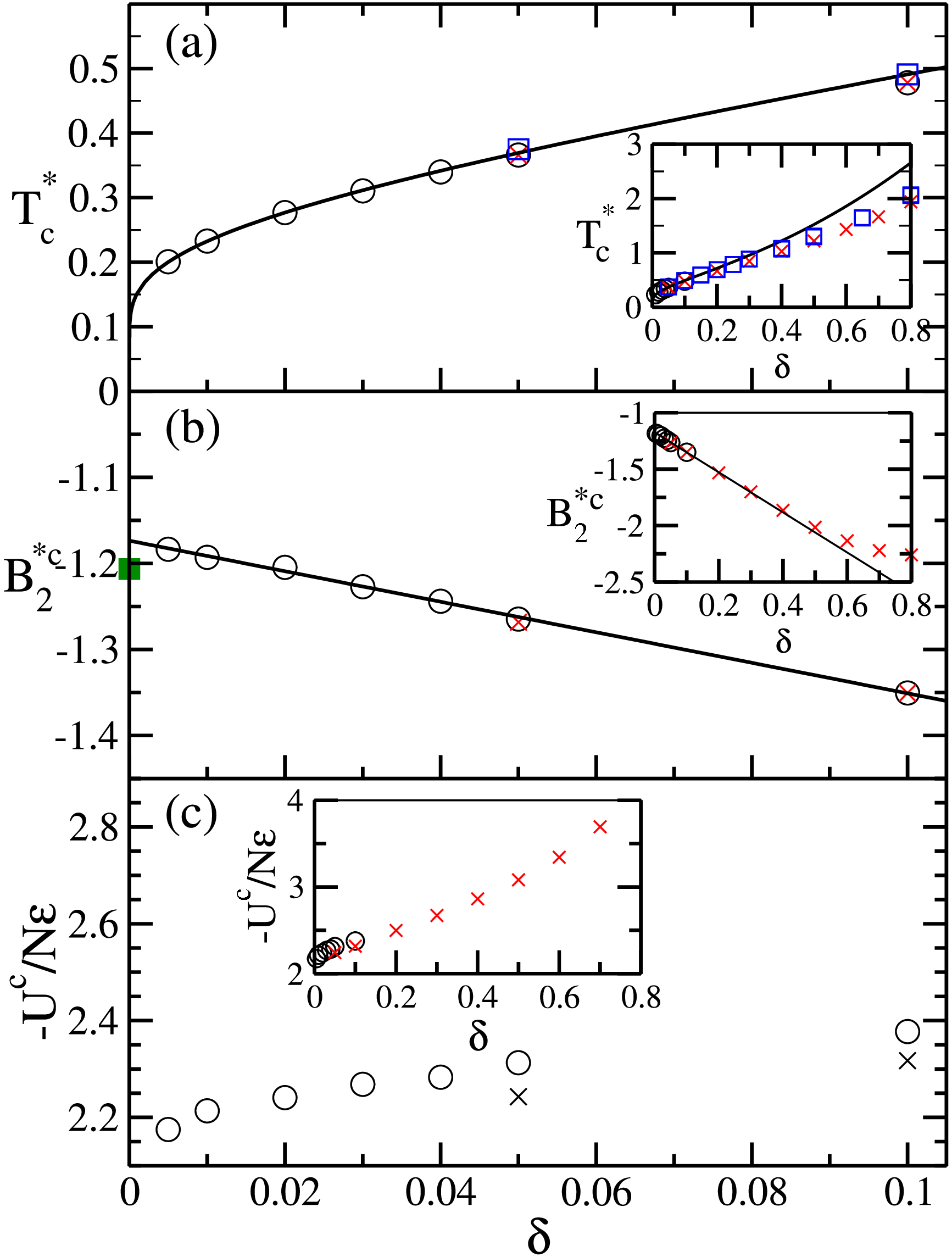}
    \caption{
    Dependence of the critical temperature (a), of the second virial coefficient (b) and of the potential energy (c) on the range of the potential $\delta$. 
    Circles and crosses label simulation data of this work with box sides 5$\sigma$ and 8$\sigma$ respectively.     The line in (a) is the theoretical prediction for the critical temperature provided by Eq.\ref{eq:Tc}.  The line in (b) corresponds to the best linear fit of the simulation data for $\delta\leq0.10$ ($-1.174-1.774\delta$).
    Open squares correspond to the  simulation data from Ref.~\cite{Lomakin1996}.  Filled squares are the AHS $B_2^{*c}$ result from Ref.~\cite{Miller2003}. 
    The inset presents  the whole range of $\delta$ values studied.
     }
    \label{fig:TcSWAHS}
 \end{figure}

Fig.-\ref{fig:TcSWAHS}-(a) and (b) 
show the $\delta$ dependence of
the critical temperature and the corresponding critical 
 second virial coefficient 
 $B_2^{*c}$. For SW, the virial can be calculated as 
\begin{equation}
B_2^*(T)=1-((\delta+1)^3-1)(e^{1/T}-1)
\end{equation}

For small $\delta$, a clear linear dependence 
of $B_2^{*c}$   is observed. 
The data are well represented by a functional form 
$B_2^{*c}(\delta)= -1.174 - 1.774~\delta $.  The extrapolated value of $B_2^{*c}$ for $\delta=0$ ($B_2^{*c}=-1.174$) is slightly higher than the value $-1.21(1)$  estimated by Miller and Frenkel for the AHS potential. The $\delta$ dependence of $B_2^{*c}$ formally violates the idea that $B_2^{*c}$ is the correct  scaling variable for collapsing the phase-diagram of different short-range attractive potentials onto a single master curve. Nevertheless, the constant $B_2^{*c}$  approximation is sufficiently good to explain the $T_c$ dependence, due to the non-linearity of the transformation (Eq.~\ref{eq:Tc}). To prove this point we show in  Fig.-\ref{fig:TcSWAHS}-(a) $T_c$ predicted according to $B_2^{*c}=-1.174$. In this representation, the assumption of constant $B_2^{*c}$ is sufficient to describe the $\delta$ dependence of $T_c$ up to $\delta=0.05$, with an error less than 1$\%$ (growing with $\delta$).

Next we compare the energy per particle of the system at the critical point in
Fig.-\ref{fig:TcSWAHS}-(c), to provide a measure of the number of contacts per particle.
This quantity also shows a linear dependence on $\delta$.

The study of the $\delta$ dependence of the critical density $\rho_c$ has received considerably less attention than the $\delta$ dependence of $T_c$. Recently Ref.\cite{Foffi2007} suggested a plausible
relation for the $\delta$ dependence of $\rho_c$ in the limit of 
small well width:
  \begin{equation}
\rho_c(\delta)=\frac{\rho_c(0)}{(1+\delta/2)^3},
\label{eq:rhoc}
\end{equation}
The relation is
based on the  hypothesis that  $\rho_c$ should be constant if measured using the average distance between two bonded particles $(1+\delta/2)\sigma$ as the unit of length.  
The relation was also supported by a potential energy landscape
interpretation of the generalized law of corresponding states\cite{foffipre} which shows that  configurations with the same Boltzmann weight are generated under an isotropic scaling (to change the inter-particle distances preserving the same bonding pattern) and a simultaneous change of both $\delta$ and $T$ such that the bond free-energy remains constant.

\begin{figure}[htbp]
    \centering
    \includegraphics[width=15 cm, clip=true]{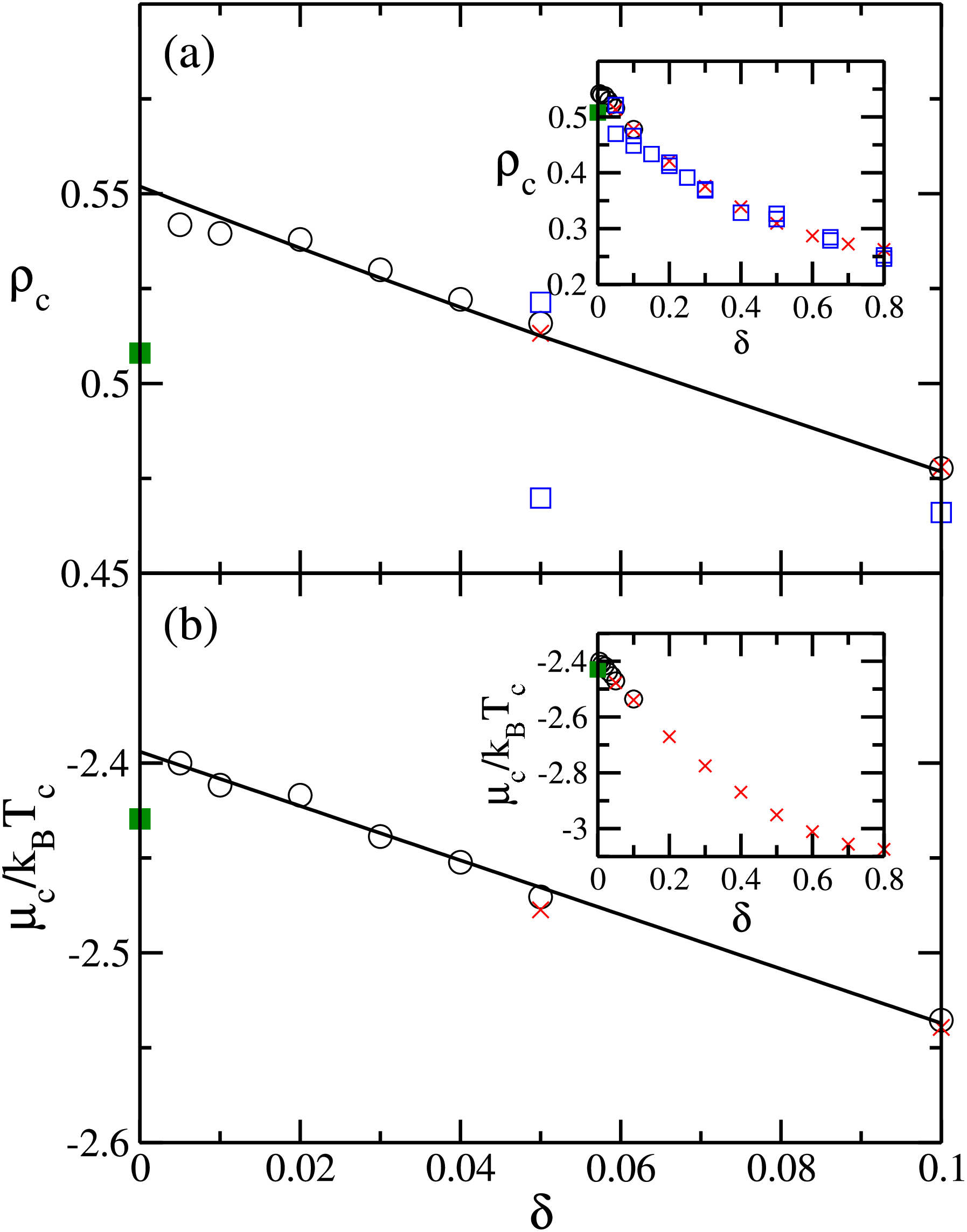}
    \caption{  Evolution of the critical density (a), and chemical potential (b) with the range of the potential $\delta$. 
    Circles and crosses correspond to  box sides $L=5\sigma$ and $L=8\sigma$ respectively. Empty squares in (a) are data from  Ref.~\cite{Lomakin1996}.
    The lines correspond to the best  fit of (a) $\rho_c$  for $\delta\leq0.10$ ($0.5519/(1+\delta/2)^3$)  and (b) $\mu_c/k_BT_c$  for $\delta\le0.10$ ($-2.394-1.431\delta$).
  Filled squares represent AHS results from Ref.~\cite{Miller2003}.    The inset presents the whole range of $\delta$ values studied in this work. }
    \label{fig:RHOcSW}
 \end{figure}

Figure~\ref{fig:RHOcSW}-(a) shows  the calculated evolution of the  critical density with the range of the interaction. It also reports
previous estimates for the same system\cite{Lomakin1996} as well as the
critical density for the AHS model from Ref.\cite{Miller2003}. As the range of the SW potential is reduced, the critical density becomes higher, since a higher local density is required to generate bonded configurations.  
Data for $\delta<0.1 \sigma$ are properly represented by
Eq.~\ref{eq:rhoc}, with a resulting fitting parameter $\rho(0)=0.552$. This value is significantly higher than the AHS critical density 
$\rho_c=0.508$ reported in Ref.-\cite{Miller2003}.

%The explanation is easier if we consider the packing fraction $\eta$. The constant parameter at criticality for different systems is not the number density but  the relative relation between the ``attractive volume" and the volume of the system $\eta^r=\frac{\pi}{6}N(\sigma+\frac{\delta\sigma}{2})^3/V$.
 
 For completeness we show in Figure~\ref{fig:RHOcSW}-(b) 
 the $\delta$ dependence of  $\mu_c/k_BT_c$, where
 $\mu_c$ is the value of the chemical potential at the critical point.
 $\mu_c/k_BT_c$ also shows a linear dependence, with an intercept
 at $-2.394$, corresponding to a critical activity $\exp(\mu_c/k_BT_c)=0.091$,
 to be compared with the corresponding value of
 $0.087$ of Miller and Frenkel.

It is known that the finite size of the system modifies the position of the critical point~\cite{Wilding1995}. A precise estimate of the critical point requires a complete finite-size scaling study to extrapolate the results to an infinite system. We have not attempted to perform such a
careful study since it would be computationally prohibitive for the small ranges studied here and we have limited
ourselves to four different box sizes ($L=5$, 6, 8 and 10) for $\delta=0.05$ only.
The results are reported in table~\ref{tab:fsize}.  As already suggested by the minor differences in the $L=5$ and $L=8$ data shown in the previous figures, no significant changes in the critical parameters are observed.
From the scatter in the data (similar to that
found by Miller and Frenkel\cite{Miller2003}) it  is possible to estimate errors in the critical parameters.
The resulting values of and errors in the critical parameters are $T_c=0.3658\pm 0.0005$, $\rho_c=0.513\pm 0.008$ and $\mu_c=0.9064\pm 0.0008$.  These values of the critical temperature and density suggest that the  Percus-Yevick  energy route \cite{Watts1971} is even better than previously thought, despite the fact that the compressibility route \cite{Baxter1968} results seem to be more often used and cited.

\begin{table}[htdp]
\caption{Critical parameters for the SW system of $\delta=0.05$ obtained with four different boxes of  side $L=5$, 6, 8, 10.}
\begin{center}
\begin{tabular}{|c|c|c|c|}
\hline
$L$ & $T_c$ & $\rho_c$ & $\mu_c$ \\
\hline
5   & 0.3660 & 0.516 & -0.9042 \\
6   & 0.3660 & 0.516 & -0.9051 \\
8   & 0.3658 & 0.513 & -0.9062 \\
10 & 0.3657 & 0.511 & -0.9069 \\   
\hline
\end{tabular}
\end{center}
\label{tab:fsize}
%\caption{XXXXXX}
\end{table}%

% \begin{figure}[htbp]
%    \centering
%    \includegraphics[width=15 cm, clip=true]{graphpMmix}
%   \end{figure}
     
% \begin{figure}[htbp]
%    \centering
%     \includegraphics[width=15 cm, clip=true]{graph3}
  %  \caption{Critical density against the interaction range $\delta$. Dots simulation results from this work, crosses simulation data from ref.-~\cite{Lomakin1996}. Filled green square, critical density for the AHS system from ref.-~\cite{Miller}. Line: critical temperature predicted from eq.-~\ref{eq:rho_c}.  }
  %  \label{fig:RHOcSW}
% \end{figure}

\section{Discussion}

Our study provides a set of values for the limiting AHS case, based on an accurate extrapolation of the critical parameters of the SW potential to 
$\delta \rightarrow 0$. These values are
 outside the error bars of Miller and Frenkel's investigation.  In particular, both the critical density
and the critical virial  appear to be higher than the previous estimates.

The special techniques  employed in the simulation of the AHS  system~\cite{Miller2003,Miller2004,kranendonkfrenkel}  only consider moves that 
make or break up to three contacts.  A particle can readily gain more than three contacts, since higher coordination states are established by a succession of such moves. Apparently, however, the constraint on the possible moves 
disfavors the formation of small nuclei of solid phases, since crystallization was extremely rarely observed with this algorithm, and then only in fluids with a very high mean reduced density (greater than 0.9). 
In the SW simulation, the transient solid-like nuclei are more readily formed (and indeed we do
occasionally observe crystallization during the simulation) suggesting that the SW simulations sample
a larger region of configuration space than that accessible with the 
AHS algorithm. This could indeed explain why the critical density extrapolated from the
SW simulations is significantly higher than that calculated previously.

\begin{figure}[htbp]
    \centering
    \includegraphics[width=15 cm, clip=true]{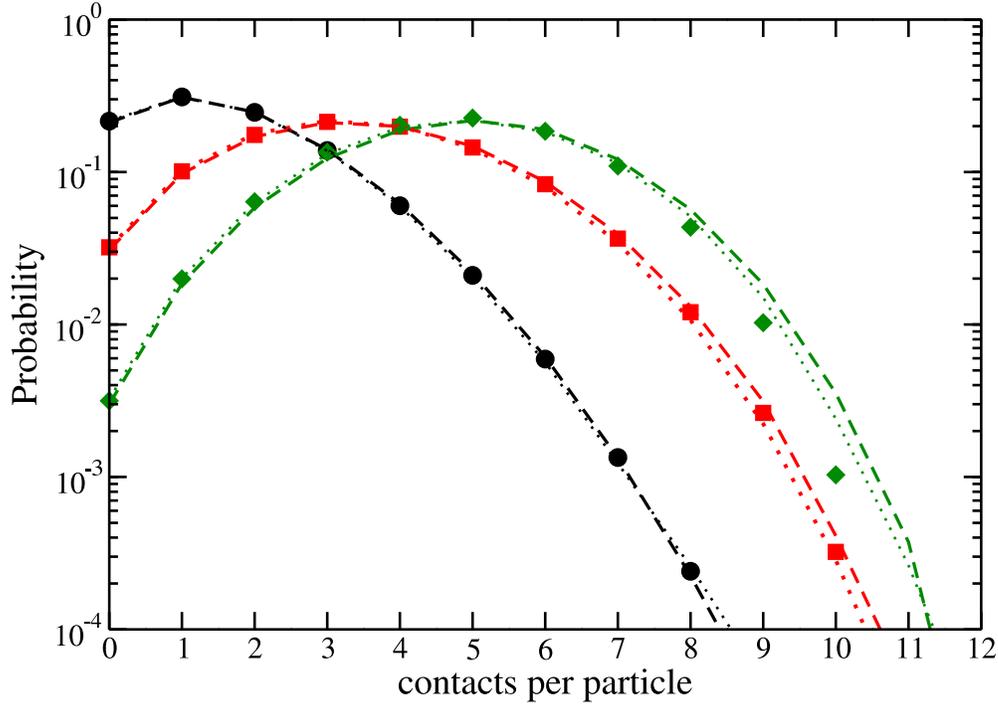}
    \caption{Probability distribution of the number of contacts per particle for the AHS system and the SW with $\delta=0.01$, for different densities. The comparison has been made at a slightly supercritical stickiness parameter $\tau$=0.120, in the case of the AHS, and the corresponding temperature provided by Eq.~\ref{eq:Tc} for the SW of $\delta=0.01$.  The simulations were performed in the NVT canonical ensemble with volume $V=512\sigma^3$. Symbols correspond to AHS simulation data, for $\rho$=0.195 (circles), 0.488 (squares), 0.781 (diamonds). Dashed lines correspond to SW simulation data and dotted lines to SW simulations at the rescaled densities (see Eq.~\ref{eq:rhoc}) 0.192, 0.481, 0.770, respectively. }
    \label{fig:cn+SW}
 \end{figure}

To support this interpretation we have compared the distribution of the number of contacts per particle
(proportional to the energy of the particle) for the AHS and a SW with $\delta=0.01$ 
at the same virial coefficient (slightly above the critical one) and same density for three different state points. The results of MC simulations in the NVT canonical ensemble, for different densities are reported in Fig.~\ref{fig:cn+SW}.   The distributions of the number of contacts per particle are coincident for low densities, but discrepancies appear as the density is increased, confirming that the algorithm used in Ref.~\cite{Miller2003} explores configurations with a somewhat smaller coordination number than does the standard MC SW simulation. Figure~\ref{fig:cn+SW} nevertheless confirms that the AHS algorithm permits the formation of high coordination states.  
To avoid any potential artifact due to the {\em a priori} unknown mapping in the density
between the AHS and the SW potential, we have also repeated the calculation at a lower density,
scaled according to Eq.~\ref{eq:rhoc}.  However, as shown in Fig.~\ref{fig:cn+SW}, even when the density is scaled to account for the different
bond distance in the AHS and SW models, at high density the disagreement between the two set of 
simulations remains.  
 
% \begin{figure}[htbp]
 %  \centering
  %  \includegraphics[width=15 cm, clip=true]{densitymodify}
  %  \caption{Critical density against the interaction range $\delta$. Dots simulation results from this work, crosses simulation data from ref.-~\cite{Lomakin1996}. Filled green square, critical density for the AHS system from ref.-~\cite{Miller}. Line: critical temperature predicted from eq.-~\ref{eq:rho_c}.  }
  %  \label{fig:RHOcSW}
 %\end{figure}

\begin{table}[htdp]
\caption{Critical point parameters for the SW fluid for  width $\delta$, resulting from simulations of  boxes of side $L=5$ (top part of the table) and 8~$\sigma$ (bottom part of the table)  respectively.}
\begin{center}
\begin{tabular}{|c|c|c|c|}
\hline
$\delta$ &$T^*_c$&$\rho_c$&$\mu_c$\\
\hline
0.005 & 0.2007 & 0.542 & -0.4817 \\
0.01 & 0.2328 & 0.540 & -0.5614 \\
0.02 & 0.2769 & 0.538 & -0.6693 \\
0.03 & 0.3106 & 0.530 & -0.7575 \\
0.04 & 0.3398 & 0.522 & -0.8333 \\
0.05 & 0.3660 & 0.516 & -0.9042 \\
0.10 & 0.4780 & 0.478 & -1.2120 \\     
\hline
0.05 & 0.3658 & 0.513 & -0.9062 \\
0.10 & 0.4780 & 0.478 & -1.2138 \\
0.20 & 0.667 & 0.421 & -1.7812 \\
0.30 & 0.847 & 0.376 & -2.3505 \\
0.40 & 1.029 & 0.339 & -2.9525 \\
0.50 & 1.220 & 0.310 & -3.6002 \\
0.60 & 1.430 & 0.287 & -4.3051 \\
0.70 & 1.665 & 0.272 & -5.0890 \\
0.80 & 1.940 & 0.263 & -5.9636 \\   
%\hline
\hline
\end{tabular}
\end{center}
\label{default}
\end{table}%

\section{Conclusions}

We have reported a simulation study aimed at evaluating the dependence of the critical parameters of the
short-range square well potential for interaction ranges approaching zero, down to $\delta=0.005 \sigma$.   The resulting values for $B_2^{*c}$ and $\beta_c \mu_c$  in the range $0.005< \delta < 0.1$ are very well described by a linear dependence on $\delta$, providing an estimate for the $\delta \rightarrow 0$ limit.  From the resulting value of $B_2^{*c}$ at $\delta=0$, it is possible to evaluate also 
the corresponding critical value of the AHS model via the relation $B_2^{*}=1-1/4\tau$.
In the same range, the critical density is well represented by the previously proposed functional form  of Eq.~\ref{eq:rhoc}.  
More precisely, our extrapolation for the AHS limit is:
\begin{eqnarray}
B_2^{*c}(\delta=0)=-1.174\pm 0.002 \\
\tau_c = 0.1150\pm 0.0001 \\
\rho_c(\delta=0) = 0.552  \pm 0.001 \\
\beta_c \mu_c(\delta=0)=-2.394  \pm 0.001
\end{eqnarray}
where the error bars are based on the spread of results from the different box sizes in this study.

Three important observations are in order:

i) Even for short-range interactions, $B_2^{*c}$ shows a slight linear dependence on $\delta$,
apparently contradicting the law of corresponding states.  While this is technically correct,
one must also remember that the small changes of $B_2^{*c}$ with $\delta$, in the range
$0< \delta < 0.05$  are not sufficient to produce any significant observable effect in the
critical temperature estimated using a constant  $B_2^{*c}$  assumption via Eq.~\ref{eq:Tc}.
This is due to the highly non-linear relationship between the two quantities, explaining the
success of the  law of corresponding states in the interpretation of simulation and experimental data of short-ranged attractive potentials.
From a more academic point of view,   the law of corresponding states
may still hold but with a scaling variable more complicated than the reduced virial coefficient itself and with a proper scaling of  the density.

%Indeed, our data suggests that for $0< \delta < 0.05$  the critical point is characterized by the same value of the potential energy and hence by the same distribution of bonds. A similar observation has also been reported in a recent study of the range dependence in patchy particles\cite{Foffi2007}.   

ii) The ``best'' critical parameters of the $\delta=0$ case are found to be different from those
reported by  Miller and Frenkel~\cite{Miller2003}. We believe this discrepancy is
related to an incomplete mapping of the configuration space which  manifests
itself in a less complete sampling of the dense region. The extreme rarity of any hint of crystallization
in the simulations of Miller and Frenkel is consistent with this
proposed explanation.

 iii) The higher critical temperature and density established by the SW  extrapolation suggest that the Percus-Yevick energy route  offers a more accurate estimate of the AHS critical point than the Percus-Yevick compressibility route.

%\newpage 

We acknowledge support from MCRTN-CT-2003-504712 and J. L. MERG-CT-2006-046453.  We thank  D. Frenkel and G. Foffi for helpful discussions.
	
%\section*{References}
%\bibliographystyle{unsrt}%unsrt,phaip,achemso
%\bibliographystyle{apsrev}
%\bibliography{SWarticle}

\begin{thebibliography}{33}
\expandafter\ifx\csname natexlab\endcsname\relax\def\natexlab#1{#1}\fi
\expandafter\ifx\csname bibnamefont\endcsname\relax
  \def\bibnamefont#1{#1}\fi
\expandafter\ifx\csname bibfnamefont\endcsname\relax
  \def\bibfnamefont#1{#1}\fi
\expandafter\ifx\csname citenamefont\endcsname\relax
  \def\citenamefont#1{#1}\fi
\expandafter\ifx\csname url\endcsname\relax
  \def\url#1{\texttt{#1}}\fi
\expandafter\ifx\csname urlprefix\endcsname\relax\def\urlprefix{URL }\fi
\providecommand{\bibinfo}[2]{#2}
\providecommand{\eprint}[2][]{\url{#2}}

\bibitem[{\citenamefont{{Hagen} et~al.}(1993)\citenamefont{{Hagen}, {Meijer},
  {Moolj}, {Frenkel}, and {Lekkerkerker}}}]{Lekkerkerker}
\bibinfo{author}{\bibfnamefont{M.~H.~J.} \bibnamefont{{Hagen}}},
  \bibinfo{author}{\bibfnamefont{E.~J.} \bibnamefont{{Meijer}}},
  \bibinfo{author}{\bibfnamefont{G.~C.~A.~M.} \bibnamefont{{Moolj}}},
  \bibinfo{author}{\bibfnamefont{D.}~\bibnamefont{{Frenkel}}},
  \bibnamefont{and} \bibinfo{author}{\bibfnamefont{H.~N.~W.}
  \bibnamefont{{Lekkerkerker}}}, \bibinfo{journal}{Nature}
  \textbf{\bibinfo{volume}{365}}, \bibinfo{pages}{425} (\bibinfo{year}{1993}).

\bibitem[{\citenamefont{{Pellicane} et~al.}(2004)\citenamefont{{Pellicane},
  {Costa}, and {Caccamo}}}]{Pellicane2004}
\bibinfo{author}{\bibfnamefont{G.}~\bibnamefont{{Pellicane}}},
  \bibinfo{author}{\bibfnamefont{D.}~\bibnamefont{{Costa}}}, \bibnamefont{and}
  \bibinfo{author}{\bibfnamefont{C.}~\bibnamefont{{Caccamo}}},
  \bibinfo{journal}{J. of Phys.: Condens. Matter}
  \textbf{\bibinfo{volume}{16}}, \bibinfo{pages}{4923} (\bibinfo{year}{2004}),
  \eprint{arXiv:cond-mat/0407335}.

\bibitem[{\citenamefont{{Rosenbaum} et~al.}(1996)\citenamefont{{Rosenbaum},
  {Zamora}, and {Zukoski}}}]{Rosenbaum1996}
\bibinfo{author}{\bibfnamefont{D.}~\bibnamefont{{Rosenbaum}}},
  \bibinfo{author}{\bibfnamefont{P.~C.} \bibnamefont{{Zamora}}},
  \bibnamefont{and} \bibinfo{author}{\bibfnamefont{C.~F.}
  \bibnamefont{{Zukoski}}}, \bibinfo{journal}{Phys. Rev. Lett. }
  \textbf{\bibinfo{volume}{76}}, \bibinfo{pages}{150} (\bibinfo{year}{1996}).

\bibitem[{\citenamefont{{Lomakin} et~al.}(1996)\citenamefont{{Lomakin},
  {Asherie}, and {Benedek}}}]{Lomakin1996}
\bibinfo{author}{\bibfnamefont{A.}~\bibnamefont{{Lomakin}}},
  \bibinfo{author}{\bibfnamefont{N.}~\bibnamefont{{Asherie}}},
  \bibnamefont{and} \bibinfo{author}{\bibfnamefont{G.~B.}
  \bibnamefont{{Benedek}}}, \bibinfo{journal}{J. Chem. Phys.}
  \textbf{\bibinfo{volume}{104}}, \bibinfo{pages}{1646} (\bibinfo{year}{1996}).

\bibitem[{\citenamefont{{Poon}}(1997)}]{Poon1997}
\bibinfo{author}{\bibfnamefont{W.~C.~K.} \bibnamefont{{Poon}}},
  \bibinfo{journal}{Phys. Rev. E} \textbf{\bibinfo{volume}{55}},
  \bibinfo{pages}{3762} (\bibinfo{year}{1997}).

\bibitem[{\citenamefont{{Likos}}(2001)}]{likos-review}
\bibinfo{author}{\bibfnamefont{C.~N.} \bibnamefont{{Likos}}},
  \bibinfo{journal}{Physics Reports} \textbf{\bibinfo{volume}{348}},
  \bibinfo{pages}{267} (\bibinfo{year}{2001}).

\bibitem[{\citenamefont{{Noro} and {Frenkel}}(2000)}]{Noro2000}
\bibinfo{author}{\bibfnamefont{M.~G.} \bibnamefont{{Noro}}} \bibnamefont{and}
  \bibinfo{author}{\bibfnamefont{D.}~\bibnamefont{{Frenkel}}},
  \bibinfo{journal}{J. Chem. Phys.}
  \textbf{\bibinfo{volume}{113}}, \bibinfo{pages}{2941} (\bibinfo{year}{2000}),
  \eprint{arXiv:cond-mat/0004033}.

\bibitem[{\citenamefont{{Foffi} et~al.}(2005)\citenamefont{{Foffi}, {Michele},
  {Sciortino}, and {Tartaglia}}}]{Foffi2005}
\bibinfo{author}{\bibfnamefont{G.}~\bibnamefont{{Foffi}}},
  \bibinfo{author}{\bibfnamefont{C.~D.} \bibnamefont{{Michele}}},
  \bibinfo{author}{\bibfnamefont{F.}~\bibnamefont{{Sciortino}}},
  \bibnamefont{and}
  \bibinfo{author}{\bibfnamefont{P.}~\bibnamefont{{Tartaglia}}},
  \bibinfo{journal}{Phys. Rev. Lett. } \textbf{\bibinfo{volume}{94}},
  \bibinfo{pages}{078301} (\bibinfo{year}{2005}),
  \eprint{arXiv:cond-mat/0410358}.

\bibitem[{\citenamefont{{Vliegenthart} and
  {Lekkerkerker}}(2000)}]{VliegenthartLekkerkerker2000}
\bibinfo{author}{\bibfnamefont{G.~A.} \bibnamefont{{Vliegenthart}}}
  \bibnamefont{and} \bibinfo{author}{\bibfnamefont{H.~N.~W.}
  \bibnamefont{{Lekkerkerker}}}, \bibinfo{journal}{J. Chem. Phys.}
  \textbf{\bibinfo{volume}{112}}, \bibinfo{pages}{5364} (\bibinfo{year}{2000}).

\bibitem[{\citenamefont{{Barker} and {Henderson}}(1967)}]{Barker1967}
\bibinfo{author}{\bibfnamefont{J.~A.} \bibnamefont{{Barker}}} \bibnamefont{and}
  \bibinfo{author}{\bibfnamefont{D.}~\bibnamefont{{Henderson}}},
  \bibinfo{journal}{J. Chem. Phys.} \textbf{\bibinfo{volume}{47}},
  \bibinfo{pages}{4714} (\bibinfo{year}{1967}).

\bibitem[{\citenamefont{{Baxter}}(1968)}]{Baxter1968}
\bibinfo{author}{\bibfnamefont{R.~J.} \bibnamefont{{Baxter}}},
  \bibinfo{journal}{J. Chem. Phys.} \textbf{\bibinfo{volume}{49}},
  \bibinfo{pages}{2770} (\bibinfo{year}{1968}).

\bibitem[{\citenamefont{{Stell}}(1991)}]{Stell1991}
\bibinfo{author}{\bibfnamefont{G.}~\bibnamefont{{Stell}}},
  \bibinfo{journal}{J. Stat. Phys.}
  \textbf{\bibinfo{volume}{63}}, \bibinfo{pages}{1203} (\bibinfo{year}{1991}).

\bibitem[{\citenamefont{{Watts} et~al.}(1971)\citenamefont{{Watts},
  {Henderson}, and J.}}]{Watts1971}
\bibinfo{author}{\bibfnamefont{R.~O.} \bibnamefont{{Watts}}},
  \bibinfo{author}{\bibfnamefont{D.}~\bibnamefont{{Henderson}}},
  \bibnamefont{and} \bibinfo{author}{\bibfnamefont{B.~R.} \bibnamefont{J.}},
  \bibinfo{journal}{Adv. Chem. Phys.} \textbf{\bibinfo{volume}{21}},
  \bibinfo{pages}{421} (\bibinfo{year}{1971}).

\bibitem[{\citenamefont{{Chen} et~al.}(1994)\citenamefont{{Chen}, {Rouch},
  {Sciortino}, and {Tartaglia}}}]{TartagliaJPCM}
\bibinfo{author}{\bibfnamefont{S.~H.} \bibnamefont{{Chen}}},
  \bibinfo{author}{\bibfnamefont{J.}~\bibnamefont{{Rouch}}},
  \bibinfo{author}{\bibfnamefont{F.}~\bibnamefont{{Sciortino}}},
  \bibnamefont{and}
  \bibinfo{author}{\bibfnamefont{P.}~\bibnamefont{{Tartaglia}}},
  \bibinfo{journal}{J. Phys.: Condens. Matter}
  \textbf{\bibinfo{volume}{6}}, \bibinfo{pages}{10855} (\bibinfo{year}{1994}).

\bibitem[{\citenamefont{{Verduin} and {Dhont}}(1995)}]{verduin}
\bibinfo{author}{\bibfnamefont{H.}~\bibnamefont{{Verduin}}} \bibnamefont{and}
  \bibinfo{author}{\bibfnamefont{J.~K.~G.} \bibnamefont{{Dhont}}},
  \bibinfo{journal}{J. Coll. Int. Sci.} \textbf{\bibinfo{volume}{172}},
  \bibinfo{pages}{425} (\bibinfo{year}{1995}).

\bibitem[{\citenamefont{{Caccamo}}(1996)}]{CaccamoPhysRep}
\bibinfo{author}{\bibfnamefont{C.}~\bibnamefont{{Caccamo}}},
  \bibinfo{journal}{Physics Reports} \textbf{\bibinfo{volume}{274}},
  \bibinfo{pages}{1} (\bibinfo{year}{1996}).

\bibitem[{\citenamefont{{Seaton} and {Glandt}}(1987)}]{seatonglandt1987}
\bibinfo{author}{\bibfnamefont{N.~A.} \bibnamefont{{Seaton}}} \bibnamefont{and}
  \bibinfo{author}{\bibfnamefont{E.~D.} \bibnamefont{{Glandt}}},
  \bibinfo{journal}{J. Chem. Phys.} \textbf{\bibinfo{volume}{87}},
  \bibinfo{pages}{1785} (\bibinfo{year}{1987}).

\bibitem[{\citenamefont{{Kranendonk} and {Frenkel}}(1988)}]{kranendonkfrenkel}
\bibinfo{author}{\bibfnamefont{W.~G.~T.} \bibnamefont{{Kranendonk}}}
  \bibnamefont{and}
  \bibinfo{author}{\bibfnamefont{D.}~\bibnamefont{{Frenkel}}},
  \bibinfo{journal}{Mol. Phys. } \textbf{\bibinfo{volume}{64}},
  \bibinfo{pages}{403} (\bibinfo{year}{1988}).

\bibitem[{\citenamefont{{Lee}}(2001)}]{Lee2001}
\bibinfo{author}{\bibfnamefont{S.~B.} \bibnamefont{{Lee}}},
  \bibinfo{journal}{J. Chem. Phys.}
  \textbf{\bibinfo{volume}{114}}, \bibinfo{pages}{2304} (\bibinfo{year}{2001}).

\bibitem[{\citenamefont{{Miller} and {Frenkel}}(2003)}]{Miller2003}
\bibinfo{author}{\bibfnamefont{M.~A.} \bibnamefont{{Miller}}} \bibnamefont{and}
  \bibinfo{author}{\bibfnamefont{D.}~\bibnamefont{{Frenkel}}},
  \bibinfo{journal}{Phys. Rev. Lett. } \textbf{\bibinfo{volume}{90}},
  \bibinfo{pages}{135702} (\bibinfo{year}{2003}),
  \eprint{arXiv:cond-mat/0301550}.

\bibitem[{\citenamefont{{Vega} et~al.}(1992)\citenamefont{{Vega}, {de Miguel},
  {Rull}, {Jackson}, and {McLure}}}]{Vega1992}
\bibinfo{author}{\bibfnamefont{L.}~\bibnamefont{{Vega}}},
  \bibinfo{author}{\bibfnamefont{E.}~\bibnamefont{{de Miguel}}},
  \bibinfo{author}{\bibfnamefont{L.~F.} \bibnamefont{{Rull}}},
  \bibinfo{author}{\bibfnamefont{G.}~\bibnamefont{{Jackson}}},
  \bibnamefont{and} \bibinfo{author}{\bibfnamefont{I.~A.}
  \bibnamefont{{McLure}}}, \bibinfo{journal}{J. Chem. Phys.}
  \textbf{\bibinfo{volume}{96}}, \bibinfo{pages}{2296} (\bibinfo{year}{1992}).

\bibitem[{\citenamefont{{Del Rio} et~al.}(2002)\citenamefont{{Del Rio},
  {Avalos}, {Espindola}, {Rull}, {Jackson}, and {Lago}}}]{DelRio2002}
\bibinfo{author}{\bibfnamefont{F.}~\bibnamefont{{Del Rio}}},
  \bibinfo{author}{\bibfnamefont{E.}~\bibnamefont{{Avalos}}},
  \bibinfo{author}{\bibfnamefont{R.}~\bibnamefont{{Espindola}}},
  \bibinfo{author}{\bibfnamefont{L.~F.} \bibnamefont{{Rull}}},
  \bibinfo{author}{\bibfnamefont{G.}~\bibnamefont{{Jackson}}},
  \bibnamefont{and} \bibinfo{author}{\bibfnamefont{S.}~\bibnamefont{{Lago}}},
  \bibinfo{journal}{Mol. Phys. } \textbf{\bibinfo{volume}{100}},
  \bibinfo{pages}{2531} (\bibinfo{year}{2002}).

\bibitem[{\citenamefont{{L{\'o}pez-Rend{\'o}n}
  et~al.}(2006)\citenamefont{{L{\'o}pez-Rend{\'o}n}, {Reyes}, and
  {Orea}}}]{LopezRendon2006}
\bibinfo{author}{\bibfnamefont{R.}~\bibnamefont{{L{\'o}pez-Rend{\'o}n}}},
  \bibinfo{author}{\bibfnamefont{Y.}~\bibnamefont{{Reyes}}}, \bibnamefont{and}
  \bibinfo{author}{\bibfnamefont{P.}~\bibnamefont{{Orea}}},
  \bibinfo{journal}{J. Chem. Phys.}
  \textbf{\bibinfo{volume}{125}}, \bibinfo{pages}{4508} (\bibinfo{year}{2006}).

\bibitem[{\citenamefont{{Pagan} and {Gunton}}(2005)}]{pagangunton}
\bibinfo{author}{\bibfnamefont{D.~L.} \bibnamefont{{Pagan}}} \bibnamefont{and}
  \bibinfo{author}{\bibfnamefont{J.~D.} \bibnamefont{{Gunton}}},
  \bibinfo{journal}{J. Chem. Phys.}
  \textbf{\bibinfo{volume}{122}}, \bibinfo{pages}{4515} (\bibinfo{year}{2005}),
  \eprint{arXiv:cond-mat/0412177}.

\bibitem[{\citenamefont{{Liu} et~al.}(2005)\citenamefont{{Liu}, {Garde}, and
  {Kumar}}}]{liukumar}
\bibinfo{author}{\bibfnamefont{H.}~\bibnamefont{{Liu}}},
  \bibinfo{author}{\bibfnamefont{S.}~\bibnamefont{{Garde}}}, \bibnamefont{and}
  \bibinfo{author}{\bibfnamefont{S.}~\bibnamefont{{Kumar}}},
  \bibinfo{journal}{J. Chem. Phys.}
  \textbf{\bibinfo{volume}{123}}, \bibinfo{pages}{4505} (\bibinfo{year}{2005}).

\bibitem[{\citenamefont{{Elliott} and {Hu}}(1999)}]{Elliot1999}
\bibinfo{author}{\bibfnamefont{J.~R.} \bibnamefont{{Elliott}}}
  \bibnamefont{and} \bibinfo{author}{\bibfnamefont{L.}~\bibnamefont{{Hu}}},
  \bibinfo{journal}{J. Chem. Phys.}
  \textbf{\bibinfo{volume}{110}}, \bibinfo{pages}{3043} (\bibinfo{year}{1999}).

\bibitem[{\citenamefont{{Chang} and {Sandler}}(1994)}]{Chang1994}
\bibinfo{author}{\bibfnamefont{J.}~\bibnamefont{{Chang}}} \bibnamefont{and}
  \bibinfo{author}{\bibfnamefont{S.~I.} \bibnamefont{{Sandler}}},
  \bibinfo{journal}{Mol. Phys. } \textbf{\bibinfo{volume}{81}},
  \bibinfo{pages}{745} (\bibinfo{year}{1994}).

\bibitem[{\citenamefont{{Bruce} and {Wilding}}(1992)}]{Bruce1992}
\bibinfo{author}{\bibfnamefont{A.~D.} \bibnamefont{{Bruce}}} \bibnamefont{and}
  \bibinfo{author}{\bibfnamefont{N.~B.} \bibnamefont{{Wilding}}},
  \bibinfo{journal}{Phys. Rev. Lett. } \textbf{\bibinfo{volume}{68}},
  \bibinfo{pages}{193} (\bibinfo{year}{1992}).

\bibitem[{\citenamefont{{Wilding}}(1997)}]{Wilding1997}
\bibinfo{author}{\bibfnamefont{N.~B.} \bibnamefont{{Wilding}}},
  \bibinfo{journal}{J. Phys.: Condens. Matter}
  \textbf{\bibinfo{volume}{9}}, \bibinfo{pages}{585} (\bibinfo{year}{1997}),
  \eprint{arXiv:cond-mat/9610133}.

\bibitem[{\citenamefont{{Foffi} and {Sciortino}}(2007)}]{Foffi2007}
\bibinfo{author}{\bibfnamefont{G.}~\bibnamefont{{Foffi}}} \bibnamefont{and}
  \bibinfo{author}{\bibfnamefont{F.}~\bibnamefont{{Sciortino}}},
  \bibinfo{journal}{J. Phys. Chem. B} \textbf{\bibinfo{volume}{111}}, 
   \bibinfo{pages}{9702-9705}
  (\bibinfo{year}{2007}).%, \eprint{0707.3114}.
  
  J. Phys. Chem. B 111, 9702-9705 (2007)

\bibitem[{\citenamefont{{Foffi} and {Sciortino}}(2006)}]{foffipre}
\bibinfo{author}{\bibfnamefont{G.}~\bibnamefont{{Foffi}}} \bibnamefont{and}
  \bibinfo{author}{\bibfnamefont{F.}~\bibnamefont{{Sciortino}}},
  \bibinfo{journal}{Phys. Rev. E} \textbf{\bibinfo{volume}{74}},
  \bibinfo{pages}{050401} (\bibinfo{year}{2006}),
  \eprint{arXiv:cond-mat/0610885}.

\bibitem[{\citenamefont{{Wilding}}(1995)}]{Wilding1995}
\bibinfo{author}{\bibfnamefont{N.~B.} \bibnamefont{{Wilding}}},
  \bibinfo{journal}{Phys. Rev. E} \textbf{\bibinfo{volume}{52}},
  \bibinfo{pages}{602} (\bibinfo{year}{1995}), \eprint{arXiv:cond-mat/9503145}.

\bibitem[{\citenamefont{{Miller} and {Frenkel}}(2004)}]{Miller2004}
\bibinfo{author}{\bibfnamefont{M.~A.} \bibnamefont{{Miller}}} \bibnamefont{and}
  \bibinfo{author}{\bibfnamefont{D.}~\bibnamefont{{Frenkel}}},
  \bibinfo{journal}{J. Phys.: Condens. Matter}
  \textbf{\bibinfo{volume}{16}}, \bibinfo{pages}{4901} (\bibinfo{year}{2004}),
  \eprint{arXiv:cond-mat/0406603}.

\end{thebibliography}

\end{document}